\documentclass[10pt]{article}
\usepackage{graphicx, setspace, amsmath, amssymb, amsfonts, amsthm, authblk, enumitem, hyperref, color, colortbl, listings}
\usepackage[linesnumbered,ruled,vlined]{algorithm2e}
\newcommand{\py}{\lstinline[language=C++]}

\definecolor{mygreen}{rgb}{0,0.6,0}
\definecolor{mygray}{rgb}{0.85,0.85,0.85}
\definecolor{mymauve}{rgb}{0.58,0,0.82}
\title{A Comparison of Dijkstra's Algorithm Using Fibonacci Heaps, Binary Heaps, and Self-Balancing Binary Trees}
\author{Rhyd Lewis}
\affil{
	School of Mathematics,\\
	Cardiff University, Cardiff, Wales.\\
	\url{LewisR9@cf.ac.uk}, \url{http://www.rhydlewis.eu}
}

\begin{document}
	\maketitle
	
	\begin{abstract}
		This paper describes the shortest path problem in weighted graphs and examines the differences in efficiency that occur when using Dijkstra's algorithm with a Fibonacci heap, binary heap, and self-balancing binary tree. Using C++ implementations of these algorithm variants, we find that the fastest method is not always the one that has the lowest asymptotic complexity. Reasons for this are discussed and backed with empirical evidence. 
	\end{abstract}
	
	\section{Introduction}
	\label{sec:intro}
	
	Dijkstra's algorithm is an efficient, exact method for finding shortest paths between vertices in edge- and arc-weighted graphs. It is particularly useful in transportation problems when we want to determine the shortest (or fastest) route between two geographic locations on a road network~\cite{Bast2016,Fu2006,Lewis2020a}. It is also applicable in areas such as telecommunication, social network analysis, arbitrage, and currency exchange~\cite{Lewis2020,Sedgewick2003}. 
	
	In this paper, we examine the changes in computational complexity and computing times that occur when using either a self-balancing binary tree, binary heap, or Fibonacci heap within Dijkstra's algorithm. As part of this work, we give an efficient C++ implementation of the algorithm and of Fibonacci heaps. Many programming languages, including C++, contain versions of self-balancing binary trees and binary heaps as part of their libraries; however, implementations of Fibonacci heaps are less common. Existing C++ implementations of Fibonacci heaps are also buggy, inefficient, and/or difficult to use. This is not the case for the custom implementation used here, which has been fully tested and evaluated.
	
	The next section formally defines the shortest path problem and surveys several algorithms for solving it. Section~\ref{sec:Dijkstra} gives a detailed description of Dijkstra's algorithm, while Section~\ref{sec:PQs} shows how the efficiency of this method can be improved through the use of priority queues. In Section~\ref{sec:implementation} we describe implementations of four variants of Dijkstra's algorithm, which are then compared and evaluated in Section~\ref{sec:evaluation}. Conclusions are drawn in Section~\ref{sec:conclusion}. 
	
	\section{Problem Definition and Existing Algorithms}
	
	Let $G = (V, A)$ be an arc-weighted, directed graph in which $V$ is a set of $n$ vertices, and $A$ is a set of $m$ arcs (directed edges). In addition, let $\Gamma(u)$ denote the set of vertices that are \emph{neighbours} of a vertex $u$. That is, $\Gamma(u) =\{v: (u,v)\in A\}$. Finally, we also define a nonnegative \emph{weight} (or length) $w(u,v)$ for each arc $(u,v)\in A$. The weight (or length) of a path is defined by the sum of the weights of its arcs. 
	
\begin{figure}[tb!]
		\centering
		\includegraphics[scale=0.4, clip= true, trim=0 0 0 0]{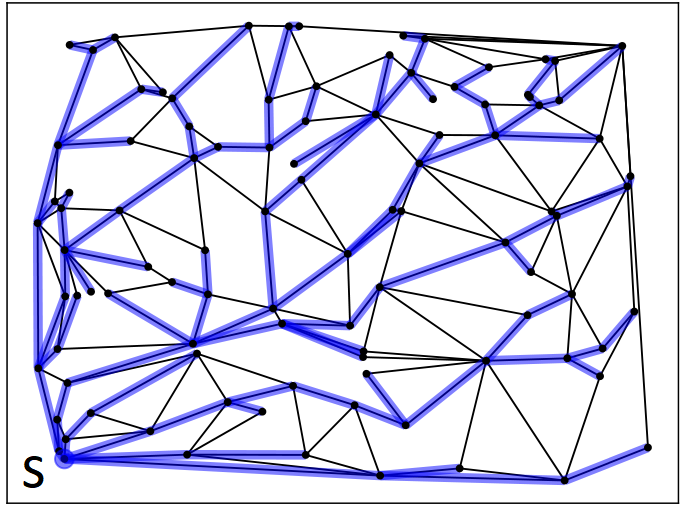}
		\caption{An example graph with $n = 100$ vertices (the black circles). In this case, arcs are drawn with straight lines, and arc weights correspond to the lengths of these lines. This particular graph is also symmetric in that $(u, v)\in A$ if and only if $(v, u)\in A$, with $w(u, v) = w(v, u)$. The highlighted arcs show a shortest path tree rooted at the vertex $s$. Because this graph is a strongly connected component, the shortest path tree is also a spanning tree.}
		\label{fig:tree}
	\end{figure}

	According to Cormen~et~al.~\cite{Cormen2000}, three problems involving shortest paths on arc-weighted graphs can be distinguished: 
	\begin{description}
		\item[The single-source single-target shortest path problem.] This involves finding the shortest path between a particular \emph{source} vertex $s$ and \emph{target} vertex $t$. In other words, we want to identify the $s$-$t$-path in $G$ whose length (weight) is minimal among all possible $s$-$t$-paths.
		\item[The single-source shortest path problem.] This involves determining the shortest path from a source $s$ to all other reachable vertices in $G$. In this sense, we are seeking a ``shortest path tree rooted at $s$''. An example of such a tree is shown in Figure~\ref{fig:tree}.
		\item[The all-pairs shortest path problem.] This involves finding the shortest path between every pair of vertices in $G$. That is, we are seeking the shortest $u$-$v$-paths for all $u,v\in V$. 
	\end{description}
	
	In this work, we will use Dijkstra’s algorithm to solve the second problem in the above list. It can, however, also be used for the other two variants. To do this with the single-source single-target shortest path problem, we simply need to halt Dijkstra's algorithm as soon as the target vertex becomes ``distinguished'' (see Section~\ref{sec:Dijkstra}). For the all-pairs shortest path problem, meanwhile, it is sufficient to execute Dijkstra's algorithm $n$ times, using each vertex $u\in V$ as the source in turn. 
	
	Several other algorithms also exist for the above three problems. In cases where graphs feature negative arc weights, a more suitable alternative is the $O(nm)$-time Bellman-Ford algorithm~\cite{Cormen2000}. Although this algorithm has a higher growth rate than that of Dijkstra's, it has the added advantage of being able to detect if a particular instance of the shortest path problem is ``ill-defined'', in that it contains negative cycles. Bellman-Ford can also be augmented with additional data structures to form Moore's algorithm~\cite{Moore1959} which, though still featuring a complexity of $O(nm)$, usually features faster run times than Bellman-Ford. 
	
	For the single-source single-target shortest path problem, other specialised algorithms exist, though none of these is known to run asymptotically faster than Dijkstra's algorithm. One well-known alternative is the A* algorithm of Hart et al.~\cite{Hart1968,Hart1972}. This is a heuristic-based variant of Dijkstra's algorithm and usually gives much faster run times in applications involving transportation networks. Algorithms for variants of the single-source single-target shortest path problem have also been proposed by Yen~\cite{Yen1971} and Bhandari~\cite{Bhandari1999}. Yen's algorithm is used to find the $k$ shortest $s$-$t$-paths, where $k$ is a user-defined parameter. The methods of Bhandari, meanwhile, are used to produce a pair of shortest $s$-$t$-paths that are either edge-disjoint and/or vertex-disjoint. 
	
	Finally, alternative algorithms are also available for the all-pairs shortest path problem. One well-known approach is the $O(n^3)$ Floyd-Warshall algorithm, which is also able to handle graphs containing negative arc weights. Another option in the presence of negative weights is the algorithm of Johnson~\cite{Johnson1977}. This operates by transforming the input graph into a second graph that has no negative weights but which maintains the same shortest-paths structure as the original. This can be achieved through a single application of the Bellman-Ford algorithm. After this, $n$ applications of Dijkstra's algorithm can then be performed. 	
	
	\section{Dijkstra's Algorithm}
	\label{sec:Dijkstra}
	
	In this section, we give a more detailed description of Dijkstra's algorithm. We also describe its underlying data structures and derive its complexity. 
	
	To produce a shortest-path tree rooted at $s$, Dijkstra's algorithm operates by maintaining a set $D$ of so-called ``distinguished vertices''. Initially, only the source vertex $s$ is considered distinguished. During execution, further vertices are then added to $D$, one at a time, until all reachable vertices have been inserted. Two other data structures are also maintained. First, a ``label'' $L(u)$ is stored for each vertex $u\in V$ in the graph. During execution, $L(u)$ stores the length of the shortest $s$-$u$-path that uses distinguished vertices only. Consequently, on termination of the algorithm, $L(u)$ gives the length of the shortest $s$-$u$-path in the graph. If a vertex $u$ has a label $L(u)=\infty$, then no $s$-$u$-path is possible. Finally, a ``predecessor'' $P(u)$ is also stored for each vertex $u\in V$. During execution, $P(u)$ stores the vertex that occurs before $u$ in the shortest $s$-$u$ path (of length $L(u)$) that uses distinguished vertices only. If a vertex $u$ has no predecessor, then $P(u)=\textsc{Null}$. On termination, these predecessor values can be used to construct the shortest paths from $s$ to all reachable vertices.
	
\begin{figure}[tb!]
		\centering
		\includegraphics[scale=0.55, clip= true, trim=0 418 130 0]{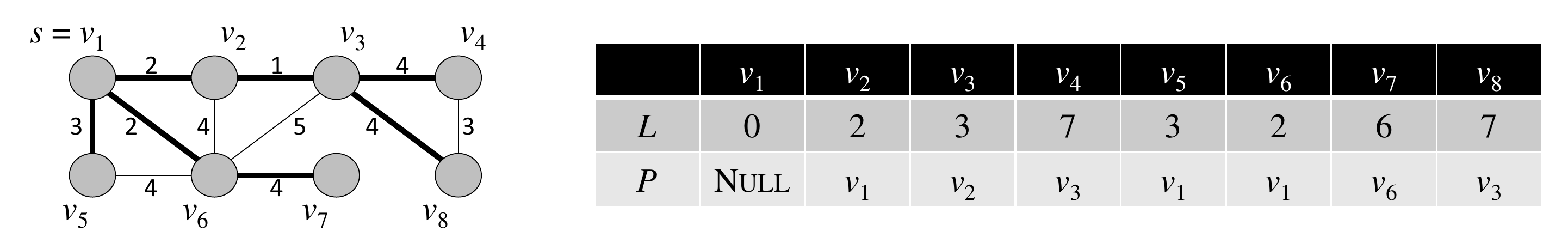}
		\caption{Example output from Algorithm~\ref{alg:DijkSimple} using the indicated eight-vertex graph and source vertex $s=v_1$. In this case, the graph is undirected. Consequently, an arc $(u,v)$ exists if and only if the arc $(v,u)$ exists. In all cases, $w(u,v) = w(v,u)$, as shown. The shortest path tree rooted at $s$ (defined by $P$) is shown by the bold lines in the graph.}
		\label{fig:eg}
	\end{figure}

	In its most basic form, Dijkstra's algorithm can now be described by just three steps. These are given in Algorithm~\ref{alg:DijkSimple}. In these steps, note that one vertex is inserted into $D$ at each iteration. This gives $O(n)$ iterations of the algorithm in total. Furthermore, within each iteration, we need to identify the vertex $u\notin D$ with the minimum label (an $O(n)$ operation) and then examine (and possibly update) the labels of all vertices $v\in\Gamma(u)$. This leads to an overall complexity of $O(m + n^2)$. Since the upper bound of $m=O(n^2)$, this complexity can be simplified to $O(n^2)$. Output from an example run of this algorithm is shown in Figure~\ref{fig:eg}. 
	
	\begin{algorithm}[tb!]
		\label{alg:DijkSimple}
		\DontPrintSemicolon
		\SetKwInOut{Input}{input}
		\SetKwInOut{Output}{output}
		\caption{Dijkstra's Algorithm (Basic Form)}
		\Input{An arc-weighted graph $G=(V,A)$ and source vertex $s\in V$} 
		\Output{A populated label array $L$ and predecessor array $P$}
		Set $L(u) =\infty$ and $P(u)=\textsc{Null}$ for all $u\in V$. Also let $D=\emptyset$, and set $L(s) = 0$.\;
		Choose a vertex $u\in V$ such that: (a) its value for $L(u)$ is minimal; (b) $L(u)$ is less than $\infty$; and (c) $u$ is not in $D$. If no such vertex exists, then end; otherwise insert $u$ into $D$ and go to Step 3.\;
		For all neighbours $v\in \Gamma(u)$ that are not in $D$, if $L(u) + w(u, v) < L(v)$ then set $L(v) = L(u) + w(u, v)$ and set $P(v) = u$. Now return to Step 2.\;
	\end{algorithm}
	
	On completion of Dijkstra's algorithm, the sequence of vertices that occurs in each shortest path starting at $s$ is stored in $P$. The shortest $s$-$u$-path (for all $u\in V$) can be constructed using the \textsc{Get-Path} procedure shown in Algorithm~\ref{alg:GetPath}. As shown, this operates by starting at $u$, and taking each preceding vertex until the source $s$ is encountered. The $s$-$u$-path is then the reverse of this sequence. For example, the shortest $s$-$v_8$-path from Figure~\ref{fig:eg} is written $\pi=(s,v_2,v_3,v_8)$. Note that, because all arc weights are assumed to be nonnegative, the paths returned by Dijkstra's algorithm will always be ``simple''. That is, they will never contain the same vertex more than once.
	
	\begin{algorithm}[tb!]
		\label{alg:GetPath}
		\DontPrintSemicolon
		\SetKwInOut{Input}{input}
		\SetKwInOut{Output}{output}
		\caption{\textsc{Get-Path}}
		\Input{The arc weighted graph $G$, predecessors $P$, source vertex $s$, and an arbitrary vertex $u$}
		\Output{A vertex sequence $\pi$ corresponding to the shortest $s$-$u$-path in $G$}
		Let $\pi = ()$, and let $v = u$\;
		\If{$P(u)\neq\textsc{Null}$}{
			\While{$v \neq s$} {
				Append $v$ to $\pi$\ and set $v=P(v)$\;
			}
			Append $s$ to $\pi$ and then reverse $\pi$\;
		}
	\end{algorithm}
	
	As mentioned earlier, Dijkstra's algorithm is also exact, meaning that it is guaranteed to determine the shortest $s$-$u$-path for all reachable vertices $u\in V$. Short proofs of this correctness can be found in several well-known textbooks such as \cite{Cormen2000} and \cite{Rosen2018}. 
		
	\section{Using Priority Queues}
	\label{sec:PQs}
	
	Although the complexity of Dijkstra's algorithm is $O(n^2)$, for sparse graphs its run times can be significantly improved by making use of a priority queue. During execution, this priority queue is used to hold the labels of all vertices that have been considered by the algorithm but that are not yet marked as distinguished. It should also allow us to quickly identify the undistinguished vertex that has the minimum label value. This ``improved'' version of Dijkstra’s algorithm is expressed in Algorithm~\ref{alg:Dijk}.
	
	\begin{algorithm}[tb!]
		\label{alg:Dijk}
		\DontPrintSemicolon
		\SetKwInOut{Input}{input}
		\SetKwInOut{Output}{output}
		\caption{\textsc{Dijkstra}}
		\Input{An arc weighted graph $G=(V,A)$, and source vertex $s\in V$} 
		\Output{A populated label array $L$ and predecessor array $P$}
		For all $u\in V$, set $L(u) = \infty$, set $D(u) =$ false, and set $P(u) =$ \textsc{Null}\;
		Set $L(s) = 0$ and insert the ordered pair $(L(s), s)$ into $Q$\;
		\While{$Q$\textup{ is not empty }}{
			Let $(L(u), u)$ be the element in $Q$ with the minimum value for $L(u)$\;
			Remove the element $(L(u), u)$ from $Q$\;
			Set $D(u) =$ true\;
			\ForEach{$v \in \Gamma(u)$ \textup{such that} $D(v) =$ \textup{false}}{
				\If{$L(u) + w(u, v) < L(v)$}{
					\If{$L(v) < \infty$}{
						Decrease the key of $(L(v), v)$ to $L(u) + w(u, v)$. That is, replace the element $(L(v), v)$ in $Q$ with the element $(L(u) + w(u, v), v)$\;
					}
					\Else{
						Insert the element $(L(u) + w(u, v), v)$ into $Q$\;
					}
					Set $L(v) = L(u) + w(u, v)$ and set $P(v) = u$
				}
			}
		}
	\end{algorithm}
	
	As shown, the \textsc{Dijkstra} procedure in Algorithm~\ref{alg:Dijk} uses four data structures, $D$, $L$, $P$ and $Q$. The first three of these contain $n$ elements and should allow direct access (e.g., by using arrays). $D$ is used to mark the distinguished vertices, while $L$ and $P$ hold the labels and predecessors of each vertex as before. In this pseudocode, the priority queue is denoted by $Q$. At each iteration, $Q$ is used to identify the element $(L(u),u)$, representing the undistinguished vertex $u$ with the minimal label value. In the remaining instructions, this element is removed from $Q$, $u$ is marked as distinguished and, if necessary, adjustments are made to the labels of undistinguished neighbours of $u$ and the corresponding entries in $Q$.
	
	The running time of \textsc{Dijkstra} now depends on the data structure used for the priority queue $Q$. Our first option here is to use a self-balancing binary tree. These are a class of binary trees that automatically keep their height logarithmic to the number of elements they contain. In C++, implementations of a self-balancing binary tree are provided by the \py{std::set} container, usually using a red-black tree~\cite{C++Set}. 
	
	A second option for $Q$ is to use a binary heap. Binary heaps are data structures that take the form of \emph{complete} binary trees. Because of this restriction, unlike self-balancing binary trees they can be implemented using an array. This allows them to be stored in contiguous memory and also means that the parent and children of any node can be determined using arithmetic on array indices, as opposed to pointers. In C++, binary heaps are provided by the \py{std::priority_queue} container~\cite{C++PQ}.
	
	Our final option for $Q$ is to use a Fibonacci heap. This data structure operates by maintaining several heap-ordered trees. It also features better amortized running times than the previous two options for some of the operations used by Dijkstra's algorithm. Note, however, that C++ does not contain a Fibonacci heap in its standard library. Instead, a custom class is required.
	
	\begin{table}[tb!]
		\begin{center}
			\begin{tabular}{lllll}
				\hline\hline
				& Identify-Minimum & Remove-Minimum & Insert         & Decrease-Key   \\ \hline
				Self-balancing binary tree & $O(1)$           & $O(1)^\dagger$ & $O(\lg n)$     & $O(\lg n)$     \\
				Binary heap                & $O(1)$           & $O(\lg n)$     & $O(\lg n)$     & $O(\lg n)$     \\
				Fibonacci heap             & $O(1)$           & $O(\lg n)^\dagger$     & $O(1)$ & $O(1)^\dagger$ \\ \hline
			\end{tabular}
		\end{center}
		\caption{Complexities of several operators using self-balancing binary trees, binary heaps, and Fibonacci heaps. Here, $n$ represents the number of elements in the data structure. All entries are worst-case run times except for those marked by $^\dagger$, which are amortised run times. Note that the Decrease-Key operation is not available in the binary heap implementation provided by \py{std::priority_queue} in C++~\cite{C++PQ}.}
		\label{table:bigO}
	\end{table}
	
	
	Table~\ref{table:bigO} considers these three alternatives for $Q$ and uses big O notation to summarise the complexities of operations relevant to Dijkstra's algorithm. Further information on how these data structures work ``under the hood'' can be found in \cite{Cormen2000}. For self-balancing binary trees, observe that the removal of the minimum element $(L(u),u)$ (Line~5 of \textsc{Dijkstra}) takes constant amortised time. At Line~10 of the algorithm, Decrease-Key operations are then carried out by finding and removing the element $(L(v),v)$ in $Q$ and inserting the new element $(L(u)+w(u,v), v)$. This process has a complexity $O(\lg n)$. Similarly, the Insert operation on Line~12 also has a complexity of $O(\lg n)$. Using a self-balancing binary tree for $Q$, therefore, leads to an overall complexity for \textsc{Dijkstra} of $O(m \lg n)$.
	
	As shown in Table~\ref{table:bigO}, the operations with binary heaps show similar complexities to self-balancing binary trees, though the removal of the minimum element is now $O(\lg n$) as opposed to constant amortised time. Note, however, that the \py{std::priority_queue} container in C++ does not feature the functionality for performing Decrease-Key operations or for removing arbitrary elements in $Q$. The use of this data structure, therefore, requires modifications to the \textsc{Dijkstra} procedure. Specifically, at Line~10, instead of replacing the element $(L(v), v)$ in $Q$ with $(L(u)+w(u,v), v)$, the latter element is now simply inserted into $Q$ alongside the former. This means that, unlike previously, a vertex $v$ can occur in several elements of $Q$. Because of this, an additional check is now required between Lines~5 and 6 of \textsc{Dijkstra} to determine if the selected vertex $u$ is already distinguished (that is, if $D(u) =$ true). If this is the case, then the remaining steps in the while-loop should not be considered, and the process should return to Line~3. Note that this modification increases the overall complexity of \textsc{Dijkstra} to $O(m \lg m$). On the other hand, binary heaps are usually seen to be faster than self-balancing binary trees because they use contiguous memory, require fewer memory allocations, and therefore tend to feature lower constant factors in their operators. The effects of this trade-off are considered in Section~\ref{sec:evaluation}. 
	
	Finally, Table~\ref{table:bigO} also shows the complexities of these operations using Fibonacci heaps. For this data structure, Insert and Decrease-Key operations both take place in constant amortised time. In particular, unlike self-balancing binary trees, the Decrease-Key operation does not involve a removal followed by an insertion; instead, it operates by identifying the correct element $(L(v), v)$ in the heap, and then lowering the first value of the element to $L(u) + w(u, v)$, modifying its position in the heap as applicable. The use of a Fibonacci heap for $Q$, therefore, leads to an overall complexity for \textsc{Dijkstra} of $O(m + n \lg n)$. This is better than both of the previous options. Despite this, however, Fibonacci heaps are often considered to be slow in practice due to their larger memory consumption and the high constant factors contained in their operators. Indeed, it is noted by Cormen~at~al.~\cite{Cormen2000} that:
	\begin{quote}
		``the constant factors and programming complexity of Fibonacci heaps makes them less desirable than ordinary binary (of $k$-ary) heaps for most applications. Thus Fibonacci heaps are predominantly of theoretical interest.''
	\end{quote}
	This claim will also be investigated further in the next section. A summary of the complexities of these algorithm variants is provided in Table~\ref{table:summary}.
	
	\begin{table}[tb!]
		\begin{center}
			\begin{tabular}{lll}
				\hline\hline
				Variant                    & Complexity    & Comments                                                  \\ \hline
				Basic form  (Section~\ref{sec:Dijkstra})                & $O(n^2)$      & Minimum label $L(u)$ found using linear search. Optimal   \\
				&               & bound for dense graphs.                                   \\
				Self-balancing binary tree & $O(m \lg n)$  & More efficient than the basic form with sparse graphs.    \\
				Binary heap                & $O(m \lg n)$  & Same complexity as previous. Note that the C++ implementation of        \\
				&               & binary heaps (std::priority\_queue) does not allow the removal of \\
				&               & arbitrary elements, so the size of the heap is $O(m)$, leading to a\\
				&               & complexity of $O(m \lg m)$.                     \\
				Fibonacci heap             & $O(m+n\lg n)$ & Lower complexity than the previous variants.              \\ \hline
			\end{tabular}
		\end{center}
		\caption{Complexities of the different variants of Dijkstra's algorithm considered in this paper. Recall that $n$ gives the number of vertices in the graph, and $m$ is the number of arcs.}
		\label{table:summary}
	\end{table}

	\section{An Implementation}
	\label{sec:implementation}
	
	In this section, we consider four C++ implementations of Dijkstra's algorithm. The first three use a self-balancing binary tree (\py{std::set}), a binary heap (\py{std:priority_queue}), and a Fibonacci heap, as described in the previous section. The fourth implements the basic form of Dijkstra's algorithm seen in Section~\ref{sec:Dijkstra}.
	
	A complete listing of our code is shown in Appendix~\ref{app:code} and can be downloaded at~\cite{SourceCode}. The bespoke \py{FibonacciHeap} class is defined on Lines~79 to 296. Lines 299 to 448 then give our four implementations of Dijkstra's algorithm: \py{dijkstraFibonacci(...)}, \py{dijkstraTree(...)}, \py{dijkstraHeap(...)}, and \py{dijkstraBasic(...)}. The \py{main()} function on Lines 464 onwards applies these methods to a small toy graph. The output is shown at the end of Appendix~\ref{app:code}.
	
	The following features in this code should be noted.
	\begin{itemize}
		\item Here, graphs are defined using the custom \py{Graph} class. Graph objects are directed and are stored using the adjacency list representation for weighted graphs~\cite{WeightedGraphs}. It is assumed that the vertices are labelled from $0$ to $n-1$ (where $n$ is the number of vertices). The algorithm will also work perfectly well on undirected graphs by ensuring that if an edge $\{u,v\}$ is present in the graph, then both of the arcs $(u,v)$ and $(v,u)$ are present in the adjacency list.
		\item All arc weights in the graph are assumed to be nonnegative integers. The code uses the inbuilt C++ constant \py{INT_MAX} to represent infinity values. Exceeding this value will result in overflow and incorrect behaviour.
		\item Each version of Dijkstra's algorithm returns two arrays (vectors): the label vector $L$ and the predecessor vector $P$.
		\item Lines 450 to 462 of the code also give a function for extracting a path from $P$. This corresponds to Algorithm~\ref{alg:GetPath} above.
	\end{itemize}
	
	In the following, all trials were performed on a 64-bit Windows 10 machine with a 3.3 GHz Pro Intel Core i5-4590 CPU and 8 GB of RAM. In our case, the code was compiled using Microsoft Visual Studio 2019 under release mode. 
	
	\section{Empirical Evaluation}
	\label{sec:evaluation}
	
\begin{figure}[tb!]
		\centering
		\includegraphics[scale=0.7, clip= true, trim=80 55 70 55]{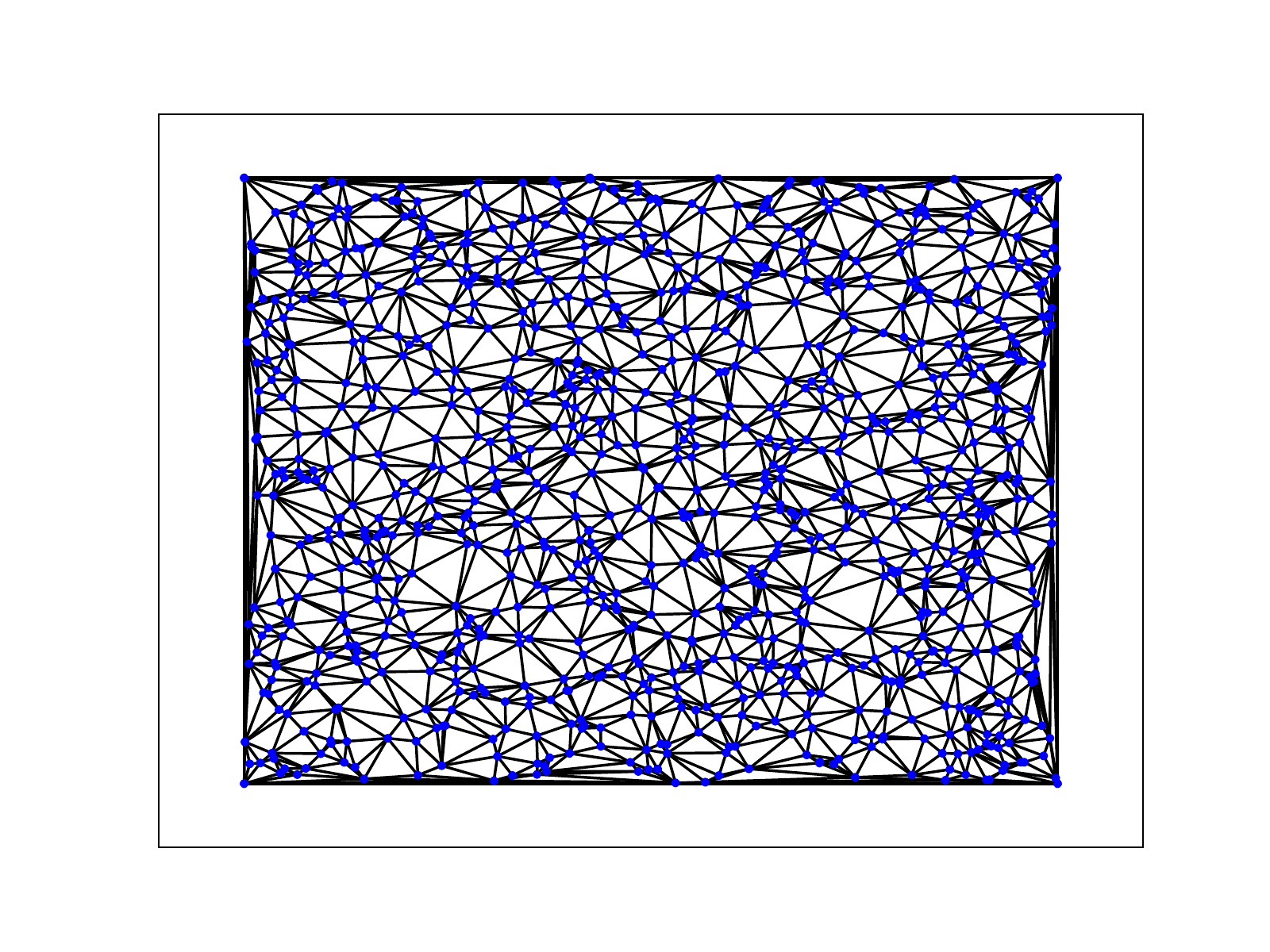}
		\caption{Example of the dense planar graphs used in our trials. This particular instance has $n=1000$ vertices and $m=5984$ arcs.}
		\label{fig:planareg}
	\end{figure}

	To assess the performance of the four variants of Dijkstra's algorithm, timed tests were carried out on two graph topologies: dense planar graphs and random graphs. Planar graphs are a type of graph that can be drawn on a plane so that no arcs intersect. Here, they were formed by randomly placing $n$ vertices into a $(10,000\times 10,000)$-unit square before generating a random Delaunay triangulation to give a graph with approximately (but not exceeding) $6n - 12$ arcs. The weight of each arc was then set to the Euclidean distance between its two endpoints, giving $w(u,v)=w(v,u)$ for all $(u,v)\in A$. An example is shown in Figure~\ref{fig:planareg}. These planar graphs can be considered similar to road networks which, as noted, are an important application area of shortest path algorithms. Our random graphs, meanwhile, were generated by creating $n$ vertices and then, for each ordered pair of vertices $u, v$, adding the arc $(u, v)$ with a probability $p$. This generation process leads to graphs with approximately $pn(n-1)$ arcs. Here, each arc was assigned a weight between $1$ and $10,000$, selected at random.
	
	\begin{figure}[tb!]
		\centering
		\includegraphics[scale=0.65, clip= true, trim=60 150 40 170]{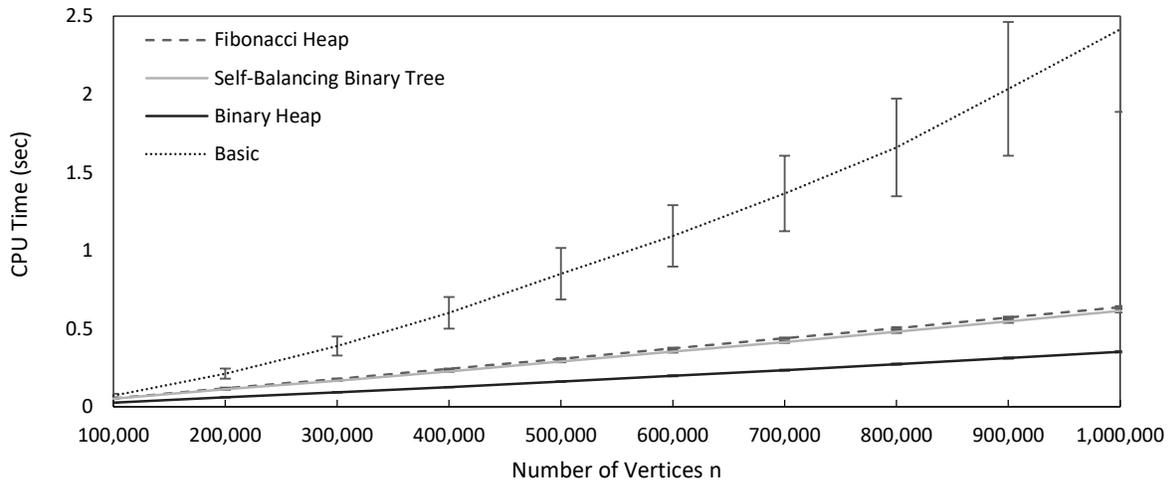}
		\caption{Execution times of the four algorithm variants with dense planar graphs. Each point on the graph is the mean taken from 100 runs. Error bars show one standard deviation on either side of the mean.}
		\label{fig:planar}
	\end{figure}
	
	Execution times of the four algorithm variants with our dense planar graphs are summarised in Figure~\ref{fig:planar} using a range of values for $n$. Across this range, we see that the use of a self-balancing binary tree gives faster run times than Fibonacci heaps, though these differences are very marginal. On the other hand, the use of a binary heap gives a noticeable improvement over these two variants, with run times dropping by approximately half. That said, each of these three algorithms can compute shortest path trees of up to a million vertices in well under a second with these graphs. As noted, the maximum number of arcs in a directed planar graph is $6n - 12$ meaning that, in these cases, $m=O(n)$. Consequently, we can consider these three variants of Dijkstra's algorithm to have a complexity of $O(n \lg n)$ here. Given that binary heaps involve fewer computational overheads than self-balancing binary trees and Fibonacci heaps, this helps to explain the superior performance of the binary heap variant in these cases. Finally, observe that the basic form of Dijkstra's algorithm shows both higher means and variances in run times compared to the other three options. This is to be expected due to the sparsity of these graphs. The gap between the basic version and the other three also widens with increases in $n$, highlighting the quadratic growth rate of the former.
	
	\begin{figure}[tb!]
		\centering
		\includegraphics[scale=0.455, clip= true, trim=60 260 60 260]{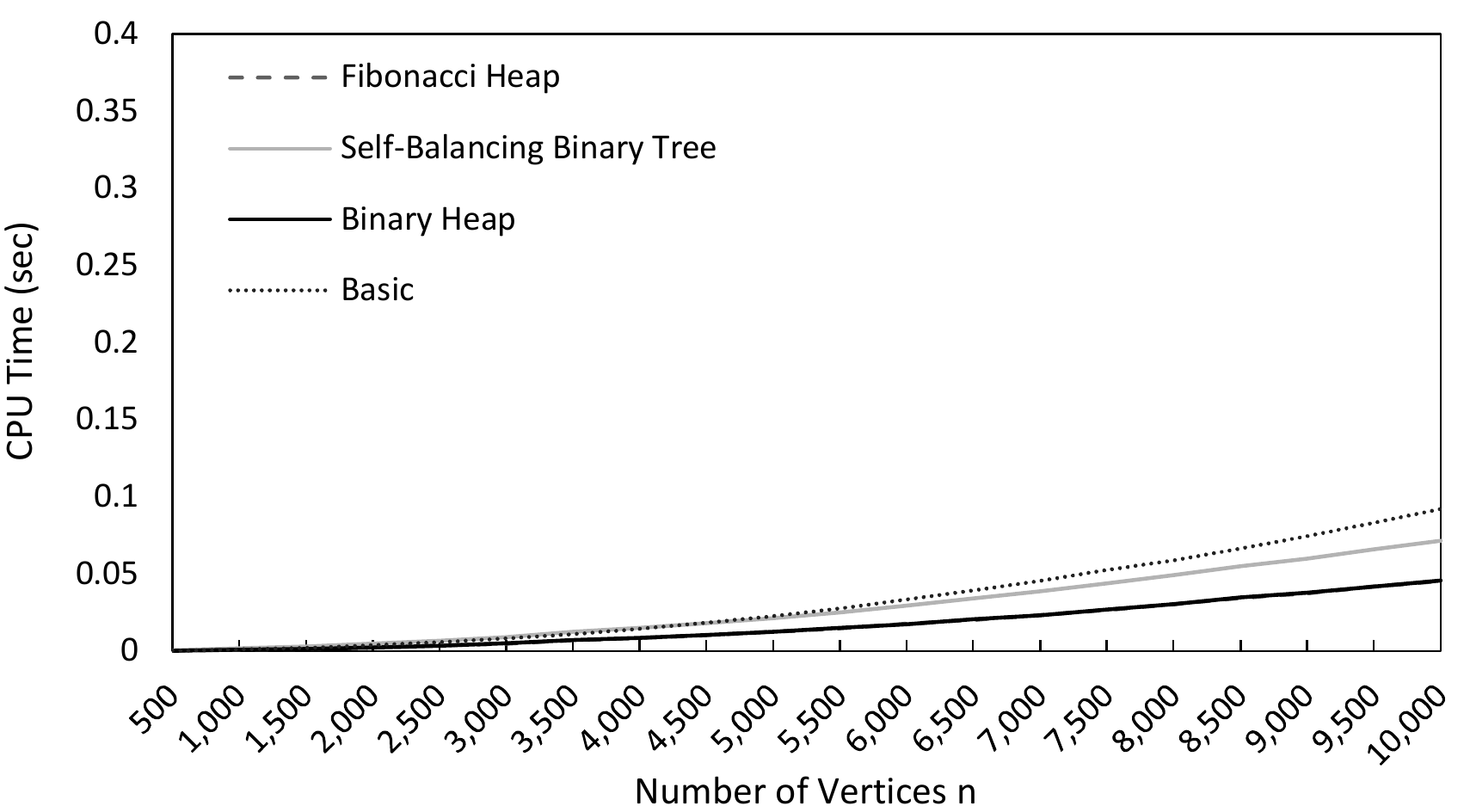}
		\includegraphics[scale=0.455, clip= true, trim=60 260 60 260]{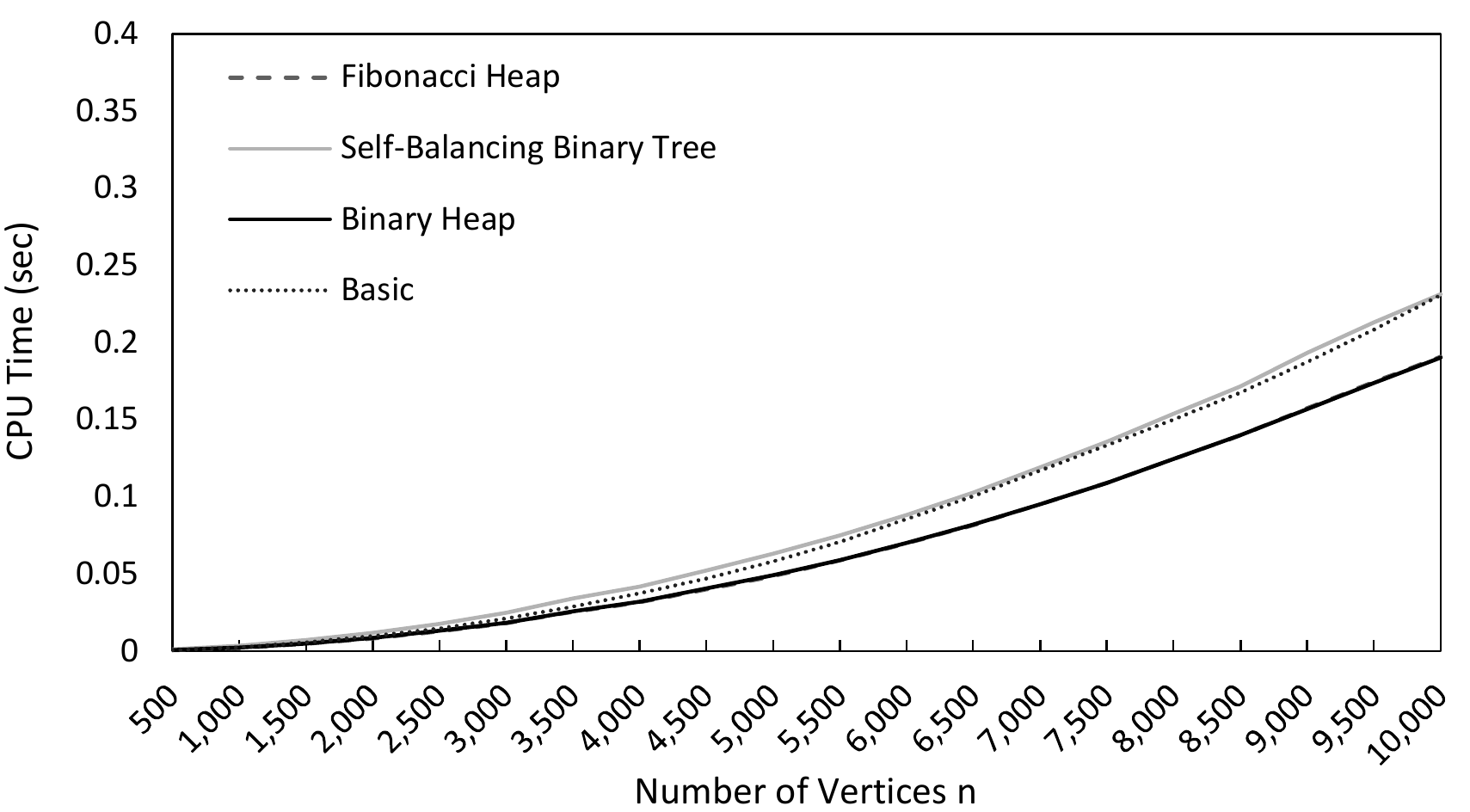}
		\includegraphics[scale=0.455, clip= true, trim=60 260 60 260]{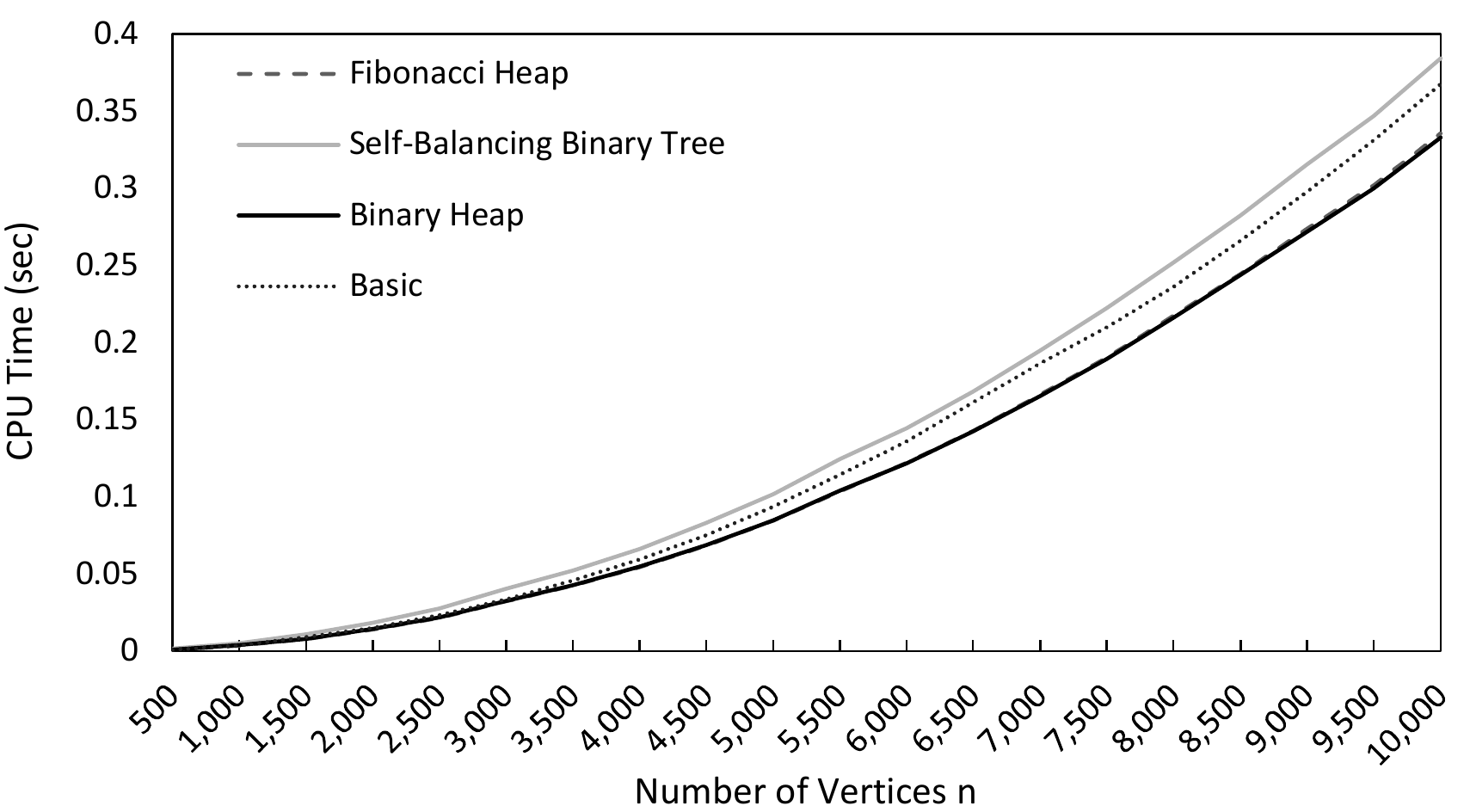}
		\caption{Execution times of the four algorithm variants with random graphs using $p= 0.1$, $0.5$ and $0.9$ respectively. Each point on the graph is the mean taken from 100 runs. Error bars are not shown here because all standard deviations were seen to be less than 0.006 seconds.}
		\label{fig:rand}
	\end{figure}
	
	The charts in Figure~\ref{fig:rand} show the results of the same experiments using random graphs with $p=0.1$, $0.5$ and $0.9$ respectively. Note that, in these cases, the number of arcs is much higher than the number of vertices; consequently, the value of $m$ is now the dominant factor in the variants using priority queues. Since $m\approx pn(n-1)$, increases to $p$ and/or $n$ therefore result in longer run times. 
	
	For these graphs, we see that the relative performance of the Fibonacci heap variant improves, with its results being almost indistinguishable from those of the binary heap. The reasons for this are that, with these denser graphs, the number of neighbours per vertex is larger. This brings a higher number of Decrease-Key and Insert operations during execution, which are more efficient with Fibonacci heaps. Despite this, however, the additional overheads required by Fibonacci heaps seem to prevent the algorithm from improving on the binary heap's run times. In cases where $n$ and/or $p$ are high, the basic version of Dijkstra's algorithm also sometimes outperforms the variant using self-balancing binary trees. This is particularly the case for graphs with $p=0.9$ where the number of arcs $m$ in these cases is close to $n^2$. 
	
	Finally, note that the largest instances considered here involve $n=10,000$ vertices, density $p=0.9$, and therefore approximately $90$ million arcs. In our runs, such graphs were seen to occupy around $900$ MB of memory, but the run times of the three variants using priority queues were still well below half a second in all cases. The times taken to load these graphs into RAM are not included in the above timings, however.
	
	\section{Conclusions}
	\label{sec:conclusion}
	
	This paper has described the shortest path problem and shown how the performance of Dijkstra's algorithm can be affected by the choice of data structure used for its priority queue. Using a C++ implementation tested over a large range of problem instances, we have seen that the best-performing algorithm does not always have the lowest complexity. Indeed, on the whole, the best performance has been seen when using Dijkstra's algorithm with a binary heap, even though its complexity of $O(m \lg m)$ is higher than the other variants. For dense graphs, however, the variant using a Fibonacci heap features very similar run times to the binary heap version. Interestingly, this binary-heap variant is also the chosen method of implementation in several open-source libraries including NetworkX~\cite{NetworkX} and GraphHopper~\cite{GraphHopper}. 	
	
	As noted, our current implementation operates by loading the entire graph into RAM before execution. It also assumes that vertices are labelled with indices from $0$ to $n-1$. In cases where these conditions are not possible, our code will need to be modified to make use of associative arrays instead of vectors. In C++ these are provided by the \py{std::map} and \py{std::unordered_map} containers.		
	
\footnotesize
\bibliography{allrefs}{}
\bibliographystyle{plain}
\normalsize

\appendix

\section{Code Listing and Example Run}
\label{app:code}

The following code, together with this paper's experimental data can be be downloaded at~\cite{SourceCode}

\begin{lstlisting}[language = C++, frame=single]
#include <iostream>
#include <climits>
#include <algorithm>
#include <vector>
#include <tuple>
#include <set>
#include <queue>
#include <time.h>

using namespace std;

const int infty = INT_MAX;

//Code for printing a vector
template<typename T>
ostream& operator<<(ostream& s, vector<T> t) {
	s << "[";
	for (size_t i = 0; i < t.size(); i++) {
		s << t[i] << (i == t.size() - 1 ? "" : ",");
	}
	return s << "] ";
}

//Struct used for each element of the adjacency list. 
struct Neighbour {
	int vertex;
	int weight;
};
//Graph class (uses adjacency list)
class Graph {
	public:
	int n; //Num. vertices
	int m; //Num. arcs
	vector<vector<Neighbour> > adj;
	Graph(int n) {
		this->n = n;
		this->m = 0;
		this->adj.resize(n, vector<Neighbour>());
	}
	~Graph() {
		this->n = 0;
		this->m = 0;
		this->adj.clear();
	}
	void addArc(int u, int v, int w) {
		this->adj[u].push_back(Neighbour{ v, w });
		this->m++;
	}
};

//Struct and comparison operators used with std::set std::priority_queue)
struct QueueItem {
	int label;
	int vertex;
};
struct minQueueItem {
	bool operator() (const QueueItem& lhs, const QueueItem& rhs) const {
		return tie(lhs.label, lhs.vertex) < tie(rhs.label, rhs.vertex);
	}
};
struct maxQueueItem {
	bool operator() (const QueueItem& lhs, const QueueItem& rhs) const {
		return tie(lhs.label, lhs.vertex) > tie(rhs.label, rhs.vertex);
	}
};

//Struct used for each Fibonacci heap node
struct FibonacciNode {
	int degree; 
	FibonacciNode* parent; 
	FibonacciNode* child; 
	FibonacciNode* left; 
	FibonacciNode* right;
	bool mark; 
	int key; 
	int nodeIndex; 
};
//Fibonacci heap class
class FibonacciHeap {
	private:
	FibonacciNode* minNode;
	int numNodes;
	vector<FibonacciNode*> degTable;
	vector<FibonacciNode*> nodePtrs;
	public:
	FibonacciHeap(int n) {
		//Constructor function
		this->numNodes = 0;
		this->minNode = NULL;
		this->degTable = {};
		this->nodePtrs.resize(n);
	}
	~FibonacciHeap() {
		//Destructor function
		this->numNodes = 0;
		this->minNode = NULL;
		this->degTable.clear();
		this->nodePtrs.clear();
	}
	int size() {
		//Number of nodes in the heap
		return this->numNodes;
	}
	bool empty() {
		//Is the heap empty?
		if (this->numNodes > 0) return false;
		else return true;
	}
	void insert(int u, int key) {
		//Insert the vertex u with the specified key (value for L(u)) into the Fibonacci heap. O(1) operation
		this->nodePtrs[u] = new FibonacciNode;
		this->nodePtrs[u]->nodeIndex = u;
		FibonacciNode* node = this->nodePtrs[u];
		node->key = key;
		node->degree = 0;
		node->parent = NULL;
		node->child = NULL;
		node->left = node;
		node->right = node;
		node->mark = false;
		FibonacciNode* minN = this->minNode;
		if (minN != NULL) {
			FibonacciNode* minLeft = minN->left;
			minN->left = node;
			node->right = minN;
			node->left = minLeft;
			minLeft->right = node;
		}
		if (minN == NULL || minN->key > node->key) {
			this->minNode = node;
		}
		this->numNodes++;
	}
	FibonacciNode* extractMin() {
		//Extract the node with the minimum key from the heap. O(log n) operation, where n is the number of nodes in the heap
		FibonacciNode* minN = this->minNode;
		if (minN != NULL) {
			int deg = minN->degree;
			FibonacciNode* currChild = minN->child;
			FibonacciNode* remChild;
			for (int i = 0; i < deg; i++) {
				remChild = currChild;
				currChild = currChild->right;
				_existingToRoot(remChild);
			}
			_removeNodeFromRoot(minN);
			this->numNodes--;
			if (this->numNodes == 0) {
				this->minNode = NULL;
			}
			else {
				this->minNode = minN->right;
				FibonacciNode* minNLeft = minN->left;
				this->minNode->left = minNLeft;
				minNLeft->right = this->minNode;
				_consolidate();
			}
		}
		return minN;
	}
	void decreaseKey(int u, int newKey) {
		//Decrease the key of the node in the Fibonacci heap that has index u. O(1) operation
		FibonacciNode* node = this->nodePtrs[u];
		if (newKey > node->key) return;
		node->key = newKey;
		if (node->parent != NULL) {
			if (node->key < node->parent->key) {
				FibonacciNode* parentNode = node->parent;
				_cut(node);
				_cascadingCut(parentNode);
			}
		}
		if (node->key < this->minNode->key) {
			this->minNode = node;
		}
	}
	private:
	//The following are private functions used by the public methods above
	void _existingToRoot(FibonacciNode* newNode) {
		FibonacciNode* minN = this->minNode;
		newNode->parent = NULL;
		newNode->mark = false;
		if (minN != NULL) {
			FibonacciNode* minLeft = minN->left;
			minN->left = newNode;
			newNode->right = minN;
			newNode->left = minLeft;
			minLeft->right = newNode;
			if (minN->key > newNode->key) {
				this->minNode = newNode;
			}
		}
		else {
			this->minNode = newNode;
			newNode->right = newNode;
			newNode->left = newNode;
		}
	}
	void _removeNodeFromRoot(FibonacciNode* node) {
		if (node->right != node) {
			node->right->left = node->left;
			node->left->right = node->right;
		}
		if (node->parent != NULL) {
			if (node->parent->degree == 1) {
				node->parent->child = NULL;
			}
			else {
				node->parent->child = node->right;
			}
			node->parent->degree--;
		}
	}
	void _cut(FibonacciNode* node) {
		_removeNodeFromRoot(node);
		_existingToRoot(node);
	}
	void _addChild(FibonacciNode* parentNode, FibonacciNode* newChildNode) {
		if (parentNode->degree == 0) {
			parentNode->child = newChildNode;
			newChildNode->right = newChildNode;
			newChildNode->left = newChildNode;
			newChildNode->parent = parentNode;
		}
		else {
			FibonacciNode* child1 = parentNode->child;
			FibonacciNode* child1Left = child1->left;
			child1->left = newChildNode;
			newChildNode->right = child1;
			newChildNode->left = child1Left;
			child1Left->right = newChildNode;
		}
		newChildNode->parent = parentNode;
		parentNode->degree++;
	}
	void _cascadingCut(FibonacciNode* node) {
		FibonacciNode* parentNode = node->parent;
		if (parentNode != NULL) {
			if (node->mark == false) {
				node->mark = true;
			}
			else {
				_cut(node);
				_cascadingCut(parentNode);
			}
		}
	}
	void _link(FibonacciNode* highNode, FibonacciNode* lowNode) {
		_removeNodeFromRoot(highNode);
		_addChild(lowNode, highNode);
		highNode->mark = false;
	}
	void _consolidate() {
		int deg, rootCnt = 0;
		if (this->numNodes > 1) {
			this->degTable.clear();
			FibonacciNode* currNode = this->minNode;
			FibonacciNode* currDeg, * currConsolNode;
			FibonacciNode* temp = this->minNode, * itNode = this->minNode;
			do {
				rootCnt++;
				itNode = itNode->right;
			} while (itNode != temp);
			for (int cnt = 0; cnt < rootCnt; cnt++) {
				currConsolNode = currNode;
				currNode = currNode->right;
				deg = currConsolNode->degree;
				while (true) {
					while (deg >= int(this->degTable.size())) {
						this->degTable.push_back(NULL);
					}
					if (this->degTable[deg] == NULL) {
						this->degTable[deg] = currConsolNode;
						break;
					}
					else {
						currDeg = this->degTable[deg];
						if (currConsolNode->key > currDeg->key) {
							swap(currConsolNode, currDeg);
						}
						if (currDeg == currConsolNode) break;
						_link(currDeg, currConsolNode);
						this->degTable[deg] = NULL;
						deg++;
					}
				}
			}
			this->minNode = NULL;
			for (size_t i = 0; i < this->degTable.size(); i++) {
				if (this->degTable[i] != NULL) {
					_existingToRoot(this->degTable[i]);
				}
			}
		}
	}
};
//End of FibonacciHeap class

tuple<vector<int>, vector<int>> dijkstraFibonacci(Graph& G, int s) {
	//Dijkstra's algorithm using a Fibonacci heap object
	int u, v, w;
	FibonacciHeap Q(G.n);
	vector<int> L(G.n), P(G.n);
	vector<bool> D(G.n);
	for (int u = 0; u < G.n; u++) {
		D[u] = false;
		L[u] = infty;
		P[u] = -1;
	}
	L[s] = 0;
	Q.insert(s, 0);
	while (!Q.empty()) {
		u = Q.extractMin()->nodeIndex;
		D[u] = true;
		for (auto& neighbour : G.adj[u]) {
			v = neighbour.vertex;
			w = neighbour.weight;
			if (D[v] == false) {
				if (L[u] + w < L[v]) {
					if (L[v] == infty) {
						Q.insert(v, L[u] + w);
					}
					else {
						Q.decreaseKey(v, L[u] + w);
					}
					L[v] = L[u] + w;
					P[v] = u;
				}
			}
		}
	}
	return make_tuple(L, P);
}

tuple<vector<int>, vector<int>> dijkstraTree(Graph& G, int s) {
	//Dijkstra's algorithm using a self-balancing binary tree (C++ set)
	int u, v, w;
	set<QueueItem, minQueueItem> Q;
	vector<int> L(G.n), P(G.n);
	vector<bool> D(G.n);
	for (u = 0; u < G.n; u++) {
		D[u] = false;
		L[u] = infty;
		P[u] = -1;
	}
	L[s] = 0;
	Q.emplace(QueueItem{ 0,s });
	while (!Q.empty()) {
		u = (*Q.begin()).vertex;
		Q.erase(*Q.begin());
		D[u] = true;
		for (auto& neighbour : G.adj[u]) {
			v = neighbour.vertex;
			w = neighbour.weight;
			if (D[v] == false) {
				if (L[u] + w < L[v]) {
					if (L[v] == infty) {
						Q.emplace(QueueItem{ L[u] + w, v });
					}
					else {
						Q.erase({ L[v], v });
						Q.emplace(QueueItem{ L[u] + w, v });
					}
					L[v] = L[u] + w;
					P[v] = u;
				}
			}
		}
	}
	return make_tuple(L, P);
}

tuple<vector<int>, vector<int>> dijkstraHeap(Graph& G, int s) {
	//Dijkstra's algorithm using a binary heap (C++ priority_queue)
	int u, v, w;
	priority_queue<QueueItem, vector<QueueItem>, maxQueueItem> Q;
	vector<int> L(G.n), P(G.n);
	vector<bool> D(G.n);
	for (u = 0; u < G.n; u++) {
		D[u] = false;
		L[u] = infty;
		P[u] = -1;
	}
	L[s] = 0;
	Q.emplace(QueueItem{ 0,s });
	while (!Q.empty()) {
		u = Q.top().vertex;
		Q.pop();
		if (D[u] != true) {
			D[u] = true;
			for (auto& neighbour : G.adj[u]) {
				v = neighbour.vertex;
				w = neighbour.weight;
				if (D[v] == false) {
					if (L[u] + w < L[v]) {
						Q.emplace(QueueItem{ L[u] + w, v });
						L[v] = L[u] + w;
						P[v] = u;
					}
				}
			}
		}
	}
	return make_tuple(L, P);
}

tuple<vector<int>, vector<int>> dijkstraBasic(Graph& G, int s) {
	//Basic Dijkstra's algorithm (O(n^2) complexity)
	int u, v, w, minL;
	size_t i, uPos;
	vector<int> L(G.n), P(G.n), Candidates;
	vector<bool> D(G.n);
	for (u = 0; u < G.n; u++) {
		D[u] = false;
		L[u] = infty;
		P[u] = -1;
	}
	L[s] = 0;
	Candidates.push_back(s);
	while (!Candidates.empty()) {
		uPos = 0;
		minL = L[Candidates[0]];
		for (i = 1; i < Candidates.size(); i++) {
			if (L[Candidates[i]] < minL) {
				minL = L[Candidates[i]];
				uPos = i;
			}
		}
		u = Candidates[uPos];
		swap(Candidates[uPos], Candidates.back());
		Candidates.pop_back();
		D[u] = true;
		for (auto& neighbour : G.adj[u]) {
			v = neighbour.vertex;
			w = neighbour.weight;
			if (D[v] == false) {
				if (L[u] + w < L[v]) {
					if (L[v] == infty) {
						Candidates.push_back(v);
					}
					L[v] = L[u] + w;
					P[v] = u;
				}
			}
		}
	}
	return make_tuple(L, P);
}

vector<int> getPath(int u, int v, vector<int>& P) {
	//Get the u-v-path specified by the predecessor vector P
	vector<int> path;
	int x = v;
	if (P[x] == -1) return path;
	while (x != u) {
		path.push_back(x);
		x = P[x];
	}
	path.push_back(u);
	reverse(path.begin(), path.end());
	return path;
}

int main() {
	//Construct a small example graph. (A directed cycle on 5 vertices here. All arcs have weight 10)
	Graph G(5);
	G.addArc(0, 1, 10);
	G.addArc(1, 2, 10);
	G.addArc(2, 3, 10);
	G.addArc(3, 4, 10);
	G.addArc(4, 0, 10);
	
	//Set the source vertex and declare some variables
	int s = 0;
	vector<int> L, P;
	
	//Execute Dijkstra's algorithm using a Fibonacci heap    
	clock_t start = clock();
	tie(L, P) = dijkstraFibonacci(G, s);
	double duration1 = ((double)clock() - start) / CLOCKS_PER_SEC;
	
	//Execute Dijkstra's algorithm using a self-balancing binary tree
	start = clock();
	tie(L, P) = dijkstraTree(G, s);
	double duration2 = ((double)clock() - start) / CLOCKS_PER_SEC;
	
	//Execute Dijkstra's algorithm using a binary heap
	start = clock();
	tie(L, P) = dijkstraHeap(G, s);
	double duration3 = ((double)clock() - start) / CLOCKS_PER_SEC;
	
	//Execute basic version of Dijkstra's algorithm
	start = clock();
	tie(L, P) = dijkstraBasic(G, s);
	double duration4 = ((double)clock() - start) / CLOCKS_PER_SEC;
	
	//Output some information
	cout << "Input graph has " << G.n << " vertices and " << G.m << " arcs\n";
	cout << "Dijkstra with Fibonacci heap took      " << duration1 << " sec.\n";
	cout << "Dijkstra with self-balancing tree took " << duration2 << " sec.\n";
	cout << "Dijkstra with binary heap took         " << duration3 << " sec.\n";
	cout << "Dijkstra (basic form) took             " << duration4 << " sec.\n";
	cout << "Shortest paths from source to each vertex are as follows:\n";
	for (int u = 0; u < G.n; u++) {
		cout << "v-" << s << " to v-" << u << ",\t";
		if (L[u] == infty) {
			cout << "len = infinity. No path exists\n";
		}
		else {
			cout << "len = " << L[u] << "\tpath = " << getPath(s, u, P) << "\n";
		}
	}
}
\end{lstlisting}

Running this code produces the following output

\begin{lstlisting}[frame=single]
Input graph has 5 vertices and 5 arcs
Dijkstra with Fibonacci heap took      1.9e-05 sec.
Dijkstra with self-balancing tree took 1.2e-05 sec.
Dijkstra with binary heap took         1.1e-05 sec.
Dijkstra (basic form) took             1.5e-05 sec.
Shortest paths from source to each vertex are as follows:
v-0 to v-0,	len = 0   path = [] 
v-0 to v-1,	len = 10  path = [0,1] 
v-0 to v-2,	len = 20  path = [0,1,2] 
v-0 to v-3,	len = 30  path = [0,1,2,3] 
v-0 to v-4,	len = 40  path = [0,1,2,3,4]
\end{lstlisting}

\end{document}